\documentclass[twocolumn,showpacs,pra,superscriptaddress,preprintnumbers]{revtex4}

\usepackage{graphicx}
\usepackage{epsfig}
\usepackage{amsfonts}
\usepackage{amsmath}
\usepackage{amssymb}
\usepackage{color}
\usepackage{overpic}

\newcommand\bea{\begin{eqnarray}}
\newcommand\eea{\end{eqnarray}}
\newcommand\beq{\begin{equation}}
\newcommand\eeq{\end{equation}}
\newcommand\bi{\bibitem}
\newcommand{\ct}{\cite}

\newcommand{\ket}[1]{|#1\rangle}

\def\non{\nonumber}
\def\dag{\dagger}

\def\de{\delta}

\def\si{\sigma}

\begin{document}
\title{Effect of double local quenches on Loschmidt echo and entanglement entropy 
of a one-dimensional quantum system}

\author{Atanu Rajak}
\email{atanu.rajak@saha.ac.in}
\affiliation{CMP Division, Saha Institute of Nuclear Physics, 1/AF Bidhannagar, Kolkata 700 064, India}
\author{Uma Divakaran}
\email{uma.divakaran@cbs.ac.in}
\affiliation{UM-DAE Center for Excellence in Basic Sciences, University of Mumbai,
Vidhyanagari Campus, Mumbai-400 098, India}

\begin{abstract}
We study the effect of two simultaneous local quenches on the evolution of Loschmidt echo and entanglement
entropy of a one dimensional transverse Ising model. 
In this work, one of the local quenches involves the connection of two spin-1/2 chains at a certain time
and the other local quench corresponds to a sudden change in the magnitude of the transverse field 
at a given site in one of the spin chains. 
We numerically calculate the dynamics associated with the Loschmidt echo and the entanglement entropy
as a result of such double quenches, and discuss various
timescales involved in this problem using the picture of quasiparticles generated as a result of such quenches.
\end{abstract}
\maketitle
\section{Introduction}
\label{intro}
Recently, quantum information theoretic measures like decoherence \ct{haroche98,zurek03,joos03}, 
entanglement and fidelity~\ct{dutta15,zanardi06,gu10,gritsev10} have become the subject of immense interest. 
Close to the quantum critical point (QCP) of a quantum many-body system,
these quantities show peculiar behaviors and hence can be regarded as a tool to detect
the QCP~\ct{dutta15}. Decoherence that signifies the loss of coherence of 
the system when it interacts with the environment is an important observable for quantum computation. 

In this context, Loschmidt echo (LE), which quantifies 
the decoherence is studied extensively  
\ct{quan06,zhang09,rossini07,sharma12,mukherjee12,nag12,rajak12}.
The LE is defined as the square of overlap of the two wave functions 
$\ket{\psi(t)}$ and $\ket{\psi_0(t)}$ evolving with two different Hamiltonians $H$ and $H_0$, respectively,
$i.e.,$ 
\beq
\mathcal{L}(t)=|\langle\psi(t)|\psi_0(t)\rangle|^2.
\label{le1}
\eeq
Initially, both the states are prepared in the 
ground state of $H_0$. 
LE provides information about how small perturbations during an evolution
can result to the decoherence of the state of the system, thus being
an important quantity for information processing and storage.
On the other hand, LE can also be used to detect the presence of a QCP by
showing a sharp dip at the QCP of the Hamiltonian
when $H$ and $H_0$ are close to each other. 

At the same time, the entanglement in quantum many-body systems,
which is the measure of quantum correlations between the two systems,
has also become a topic of intensive research interest for last several 
years~\ct{osborne02,osterloh02,vidal03,korepin04,calabrese04,refael04,igloi08,song10,dubail11}. 
One of the quantities to measure the entanglement between the two subsystems
is the von Neumann entropy~\ct{holzhey94,amico08,calabrese09,eiser10}. Consider a bipartite system 
divided into two subsystems $A$ and $B$ of length $L_A$ and $L_B$ with total length $L=L_A+L_B$. 
If the whole system is in a quantum pure state 
$|\psi\rangle$ with density matrix $\rho=|\psi\rangle\langle\psi|$, then the von Neumann entropy $S$ of 
system A with reduced density matrix $\rho_A={\rm Tr}_B(\rho)$ is defined as 
\beq
S=-{\rm Tr}(\rho_A\log\rho_A).
\label{ee1}
\eeq
The entanglement entropy (EE) or $S$ increases with increasing quantum correlations (entanglement) 
between the two subsystems. EE exhibits distinct scaling relations at and close 
to a quantum critical point with the shortest length scale of the system. 
For a critical spin chain with periodic boundary conditions where the subsystem has two boundary points,
the entanglement entropy scales as $S=\frac{c}{3}\log L_A$, where 
$c$ is a universal quantity and given by the central charge of the conformal field theory~\cite{holzhey94,vidal03,calabrese04}. 
On the other hand, away from the critical point where the correlation length $\xi\ll L_A$, 
entanglement entropy is given by $S=\frac{c}{3}\log \xi$. The above scaling relations are valid 
for a one-dimensional homogeneous system. 
It is found that some modifications are 
required in the scaling relations of EE when the system is inhomogeneous. Interestingly, 
in this case also the scaling relations remain same as the homogeneous case with 
a changed prefactor $c_{\rm eff}$ which is called the effective central 
charge~\cite{refael04,laflorencie05,chiara06,refael07,bonesteel07,igloi07,igloi071}.

With our understanding of behavior of EE in an equilibrium system getting better, 
a considerable amount of focus is also given to EE in systems out of equilibrium 
\ct{calabrese05,eisler07,calabrese07,eisler08,igloi09,hsu09,cardy11,stephan11,igloi12}.
The experimental demonstration of such non-equilibrium dynamics
using optical lattices \cite{bloch08} also contributes to the sudden upsurge in
studies related to decoherence and entanglement in out of equilibrium systems.
One of the ways of generating such a non-equilibrium dynamics
is a sudden quench. A sudden quench in the system can be performed locally or globally. 
In a global quench, a parameter of the Hamiltonian is changed suddenly 
at all the sites resulting to a non-equilibrium dynamics. 
In this process, EE generally shows a linear increase in time $t$ up to some time $t_0$~\cite{calabrese05}. 
The local quench is defined as a local change of a 
parameter of the Hamiltonian. For example, the entanglement entropy between 
two critical subsystems $A$ and $B$ of a homogeneous one-dimensional chain which are disconnected 
for $t< 0$ and connected at $t=0$ increases as $S=\frac{2c}{3}\log t$ for $t\ll L$~\cite{eisler07,calabrese07,stephan11}; 
here the final chain is periodic. On the other hand, if the final chain is open, the factor $2$ 
in the expression of $S$ is not present.
Such studies are important in the context of information propagation through a quantum many body system.

In this paper, we consider two independent transverse field Ising spin chains
in the ferromagnetic or critical phase.
For the times $t<0$, the two spin 
chains which are in their respective ground states, 
are disconnected and are later joined at $t=0$ ($J$-quenching). 
Simultaneously, we change the magnitude of the transverse field at a single 
site of one of the chains ($h$-quenching).
This results to a non-equilibrium evolution of the state of the system. 
If we consider only $J$-quenching of a critical Ising chain, both 
the quantities, namely, LE and EE show periodic time evolution \ct{stephan11}. 
However, in this paper, we address the effect 
of $h$-quenching along with the $J-$quenching in critical and off-critical systems
which results to a non-trivial evolution of LE and EE when compared to the 
no $h$-quenching case. We shall try to understand these results
using the picture of quasiparticles generated due to both the local quenches.
Our most significant observation here is the reflection of 
the quasiparticles at the site of $h$-quenching. We find some interesting 
time scales in the evolution of both LE and EE when the two quenches are performed
simultaneously. These timescales can be explained successfully using the reflection 
picture of the quasiparticles generated. 
We have also argued qualitatively how the evolution of EE 
changes as the total system moves from deep ferromagnetic phase to critical one.

The organization of the paper is as follows: we discuss the model studied in this paper
along with briefly mentioning the numerical techniques in section \ref{model}
followed by a discussion on semiclasscial theory of quasiparticles generated in section \ref{semi_cla}. 
In section \ref{le_ee_crit}, we present the results of LE and EE in the critical region for various
geometries whereas we present the results for the ferromagnetic region in section \ref{le_ee_ferro}.
A comparison between the two quenches studied in this paper is made in section \ref{sec_comparison}
using the semiclassical picture. We conclude this paper with our main results in 
section \ref{discussion}. We have also added two appendices at the end of the paper
outlining the numerical techniques used in this paper.

\section{Model}
\subsection{Exact diagonalization}
\label{model}
The Hamiltonian we consider here is that of a one-dimensional Ising chain 
in a transverse field given by
\beq
H~=~-\sum_{n}(J_n\sigma_n^x\sigma_{n+1}^x~+~h_n\sigma_n^z),
\label{ham1}
\eeq
where $h_n$ and $J_n$ are the site dependent transverse magnetic fields and cooperative interactions, 
respectively, and $\si^x_n$ and $\si^z_n$ are standard Pauli matrices at the lattice site $n$. 
For the homogeneous case ($h_n=h$ and $J_n=J$),
the model in Eq.~(\ref{ham1}) has a QCP at $J=h$ separating ferromagnetic and 
quantum paramagnetic phases.
Using Jordan-Wigner transformations followed by Fourier transformation for a homogeneous and periodic chain, 
the energy spectrum for the Hamiltonian (\ref{ham1}) is obtained as \cite{lieb61,pfeuty70}
\beq
\varepsilon_q=\pm2J\sqrt{(h+\cos q)^2+\sin^2q},
\label{spectrum1}
\eeq
where $q$ is the momentum which takes discrete values given by $q=2\pi m/L$ with 
$m=0\cdots L-1$ for a finite system of length $L$.

On the other hand, such homogeneous systems are very rare in nature. One atleast finds some local
defects, no matter how pure the material is.
The general method adopted to study systems which are not homogeneous is outlined below, which we also
use in this paper.
Following Jordan-Wigner transformation, the Hamiltonian in Eq.~(\ref{ham1}) 
can be described by a quadratic form in terms of 
spinless fermions $c_i$ and $c_i^{\dagger}$ \cite{lieb61}
\beq
H = \sum_{i,j} \left[ c_i^\dagger A_{i,j} c_j +
\frac{1}{2} ( c_i^\dagger B_{i,j} c_j^\dagger + \mathrm{h.c.}) \right].
\label{ham_quadratic1}
\eeq
Here, $\mathbf{A}$ is a symmetric matrix due to hermicity of $H$
and $\mathbf{B}$ is an antisymmetric matrix which follows from the 
anticommutation rules of $c_i$'s. 
The elements of these matrices thus obtained are:
\begin{eqnarray}
A_{i,j}&=&-(J_i\de_{j,i+1}+J_j\de_{i,j+1})-2h_i\de_{i,j},\cr
B_{i,j}&=&-(J_i\de_{j,i+1}-J_j\de_{i,j+1}).
\label{eq_AB}
\end{eqnarray}
The above Hamiltonian can be diagonalized in terms of
the normal mode spinless Fermi operators $\eta_k$ given by the relation~\ct{lieb61}. 
\bea
\eta_k=\sum_i(g_k(i) c_i + h_k(i) c_i^{\dagger}),
\label{eq_gh}
\eea
where $g_k(i)$ and $h_k(i)$ are real numbers. In terms of these operators the Hamiltonian takes 
the diagonal form,
\beq
H = \sum_k \Lambda_k \left( \eta_k^\dagger \eta_k - \frac{1}{2} \right),
\eeq
with $\Lambda_k$ being the energy of different fermionic modes with index $k$. 
These $\Lambda_k's$ are given by the solutions of 
the eigenvalue equations,
\bea
(\mathbf{A}-\mathbf{B})(\mathbf{A}+\mathbf{B})\mathbf{\Phi}_k&=&\Lambda_k^2 \mathbf{\Phi}_k \cr
(\mathbf{A}+\mathbf{B})(\mathbf{A}-\mathbf{B})\mathbf{\Psi}_k&=&\Lambda_k^2 \mathbf{\Psi}_k\;.
\label{eigen_eqs}
\eea
It can be shown that the elements of the eigenvectors are related to $\mathbf{g}$ and $\mathbf{h}$
matrices used to diagonalize the Hamiltonian as follows:
$\Phi_k(i)=g_k(i)+h_k(i)$ and $\Psi_k(i)=g_k(i)-h_k(i)$. 
We calculate LE and EE using $\Phi$, $\Psi$, $\mathbf{g}$ and $\mathbf{h}$
as discussed in appendices \ref{appendixa}, and \ref{appendixb}.

\subsection{Semiclassical theory of quasiparticles}
\label{semi_cla}

When a system at zero temperature is taken away from its ground state 
by applying some perturbation, the state of the system undergoes a 
non-equilibrium evolution with respect to the final Hamiltonian. 
The initial state, which now is an excited state,
is a source of quasiparticles (QPs) corresponding to the final Hamiltonian.
Recently, such non-equilibrium dynamics have been studied using a semiclassical picture
of quasiparticles generated for global \ct{rieger11,blass12} and local quenches \ct{divakaran11}, 
where excellent agreement between the numerics and 
the semiclassical theory were obtained. We now briefly describe this theory of
quasiparticles generated which shall be used to explain
the various timescales observed in our numerical calculations.
For global quenches from $h=0$ to a very small $h$ value, it can be shown that
these quasiparticles are wavepackets of low-lying excitations and discussed in details
in Ref. \ct{rieger11}.
Due to the conservation of momentum, quasiparticles of a given momentum 
are always produced in pairs, the group velocity 
$v_g(k)(=|\partial \varepsilon_k/\partial k|)$ of them being equal and opposite to each other.
As discussed in Ref. \ct{rieger11}, 
these quasiparticles in the small $h$ limit can be considered as 
classical particles (sharply defined QPs) which when crosses a 
site, simply flips the spin at that site. Though this picture is discussed for a very specific 
quench (a small quench),
it has been verified for stronger quenches and also for quenches in the paramagnetic phase with slight 
modifications.
It is also argued that these quasiparticles are no longer point particles, but are extended objects
as the critical point is approached due to large correlation length. 
In the following sections, we shall try to explain our numerical results 
atleast qualitatively with this point like picture of 
quasiparticles for spin chains in the ferromagnetic as well as
in the critical region.

\section{Loschmidt echo and Entanglement entropy for critical chain}
\label{le_ee_crit}
We first study double quenches for a critical chain where
already some work has been done in Ref. \cite{stephan11, calabrese07}
for local J-quenches.
As discussed before, we consider simultaneous application of two types of 
local perturbations to the system and study the time 
evolution of Loschmidt echo (LE) and von Neumann entanglement entropy (EE)
as a result of such quenches. 
Initially the spin chain is prepared in the ground state of $H=H_1$~+~$H_2$ 
where $H_1$ and $H_2$ are the Hamiltonians of
two decoupled  homogeneous Ising chains of length $L_1$ and $L_2$, respectively,
with open boundary conditions ($J_{L_1}=J_{L_2}=0$).
Two simultaneous quenches are performed at $t=0$, namely, 
(i) the two spin-1/2 chains are suddenly connected together 
resulting to a chain of total length $L=L_1+L_2$, and, 
(ii) the transverse field at a particular site $L'$ belonging to either 
the chain $1$ or $2$ is changed from $h$ to $h+\de$.
The system then evolves with the final Hamiltonian 
\beq
H_f=H_1+H_2+H_{12}^I-\de\sigma^z_{L'},
\label{ham2}
\eeq
where $H_{12}^I$ defines the connection between the two spin chains of 
length $L_1$ and $L_2$ and is of the form $J\sigma^x_{L_1}\sigma^x_{L_1+1}$.
At the same time, the term $-\de\sigma^z_{L'}$ 
in the Hamiltonian corresponds to the $h-$quenching which changes the
magnitude of transverse field at site $L'$ from $h$ to $h+\de$.
We incorporate these quenches numerically by considering
a single spin chain of total length $L=(L_1+L_2)$ with the first $L_1$
spins forming the system 1 and the remaining system 2. They are disconnected at $t<0$
by putting $J_{L_1}=0$ which at $t=0$ is then increased to $J$, also the interaction
strength at all the other sites. For all our calculations, we have set $J=1$. 
The details of the numerical calculations for LE and EE are outlined in the appendix, 
see also Refs.~[\onlinecite{rossini07,igloi09}].

As we switch on the two local perturbations discussed above,
there is a local increase in energy of the system
at the site of local perturbations \ct{calabrese07,divakaran11}. 
These sites then become the source of quasiparticle production.
Henceforth, we shall call the quasiparticles created due to the $h$-quenching at $L'$
 as $QP^1$ , and the corresponding left and right moving
quasiparticles as $QP^1_L$ and $QP^1_R$, respectively. Similarly, the left and right moving quasipartciles
created at the site $L_1$ of $J-$quenching shall be called as $QP^2_L$ and $QP^2_R$.
Below, we present our results for the evolution of LE and EE for various cases or geometry
and discuss these results in the light of quasiparticles propagating in the system.
\begin{figure}[ht]
\begin{center}
\begin{overpic}[width=7.5cm]{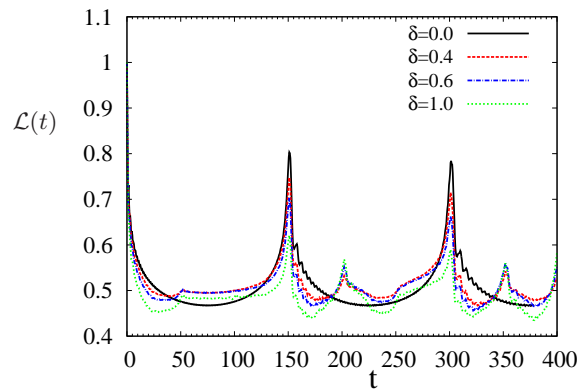} 
\put(-4,45){$\mathcal{L}(t)$}
\end{overpic}
\end{center}
\caption{The plot shows LE as a function of time for different values of $\de$
with $J-$quench at $L_1=L/2$ and the $h-$quench at site $L'=L/3$. 
The transverse field at $L'$ is changed from $1$ to $1+\de$.
For the $J$ quenching alone (i.e., $\de=0$ case), the LE shows peak at $t_3=L/v_{\text{max}}=150$ and $T=2t_3$ 
where $v_{\text{max}}=2$ and $L=300$. 
By applying two local perturbations simultaneously at time $t=0$, 
we observe a small peak at $t'=50$ and comparatively a stronger peak at $t''=200$. 
We also note small fluctuations near $t_1=100$ which is more clearly seen
for $\delta=1$ curve.}
\label{le_fig1}
\end{figure}

\subsection{LE and EE for $L_1=L_2$}
\label{sec_l1eql2}
In this section, we consider $J-$quenching at $L_1=L_2=L/2$ 
and the $h$-quenching at some site $L'$ of the total spin chain of length $L$ at time $t=0$.
Let us first discuss the dynamics of LE. 
In general, for the better readability of the paper, we shall assume $L'<L_1\leq L_A$ 
(explained later in the context of EE) through out the paper
but the case with $L'>L_1$ is also presented and discussed in the caption of various figures.
Due to the propagation of the generated quasiparticles, 
we expect four time scales which are the times of come back of the 
quasi-particles at the source point after getting reflected 
from boundaries of the chain, thus removing the effect of
their dynamics. When this happens, the overlap in the definition of LE
increases and shows a peak. 
These time scales are determined by the fastest moving QPs with 
maximal group velocity $v_{\text{max}}=\text{max}_k~v_g(k)$ and are given by 
$t_1=(2L')/v_{\text{max}}$ (time of come back of $QP^1_L$), 
$t_2=2(L-L')/v_{\text{max}}$ (time of come back of $QP^1_R$), 
$t_3=(2L_1)/v_{\text{max}}$ (time of come back of $QP^2_L$) 
and $t_4=(2L_2)/v_{\text{max}}$ (time of come back of $QP^2_R$). 
With $L_1=L_2$, the values of $t_3$ and $t_4$ are equal, but it is not the case in general.
Other than the above mentioned obvious time scales, 
two more time scales are observed numerically, $t'$ given by $2|L_1-L'|/v_{\text{max}}$, which appears to be the time
taken by $QP^2_L$ (for $L_1>L'$) to reach $L'$ and get partially 
reflected at $L'$ where it finds a change in potential
from $h$ to $h+\delta$. 
The second time scale is same as $t_2$, but it is due to $QP^2$ and is given by $t''=2(L-L')/v_{\text{max}}$,
which is the time taken by the $QP^2_L$ ($QP^2_R$) to get reflected at $L'$ (right boundary) 
and come back to $L_1$ after getting fully (partially) reflected at the right boundary ($L'$).
It is to be noted that $v_{\max}$ for the homogeneous transverse Ising model
with elements as defined in Eq. \ref{eq_AB} is $2$ at the critical point and also
in the paramagnetic phase. We shall use the same value of $v_{\max}$ in our case also 
since the numerically obtained value of $v_{max}$ by differentiating the eigenvalues
is also close to 2. 

\begin{figure}[h]
\begin{center}
\begin{overpic}[width=7.5cm]{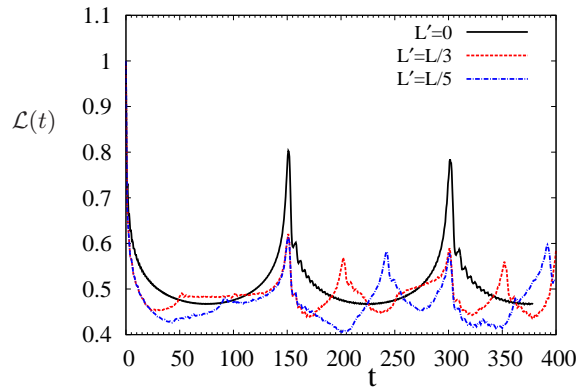}
\put(-4,45){$\mathcal{L}(t)$}
\end{overpic}
\end{center}
\caption{The plot shows LE as a function of time when $h$-quenching of strength $\de=1.0$ is performed 
at different sites $L'$ of the total chain with $J$-quenching fixed at $L_1=L/2$. Here $L'=0$ 
corresponds to the case of $J$-quenching alone where $t_3=t_4=150$.
For $L'=L/3$ ,$t'=50$ and $t''=200$ 
whereas $t'=90$ and $t''=240$ for $L'=L/5$.
All these timescales are clearly seen in the above figure.
We also observe small perturbations at $t_1$ which is not very clear.}
\label{le_fig2}
\end{figure}

\begin{figure}[h]
\begin{center}
\includegraphics[width=7.5cm]{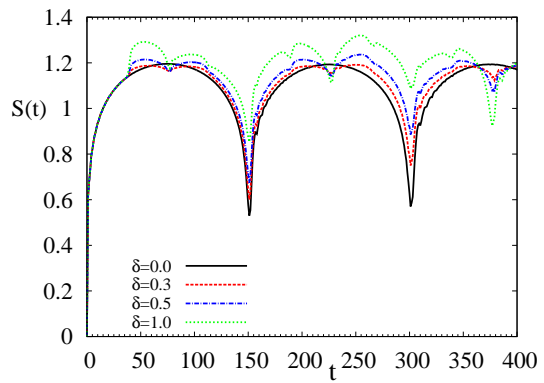}
\end{center}
\caption{Entanglement entropy as a function of time after a local $J$-quench in the 
middle of the chain ($L_1=L/2$) along with $h$-quenching at $L'=L/4$ for different 
interaction strengths $\de$. Here, we consider $L_A=L_1=L/2$ as the subsystem with total system size $L=300$. 
The time scales $t'/2$, $t'$ and $t''$ can be seen in this figure which
agrees well with our explanations. In this case, $t'=75$ and $t''=225$.
Another time scale observed
is around $t=187$ which is when $QP^1_R$ enters system A resulting to an
increase in EE as $QP^1_L$ is in system B during that time.
Note that the effect of $QP^1$ is very small compared to $QP^2$.
}
\label{ee_fig1a}
\end{figure}

\begin{figure}[h]
\begin{center}
\includegraphics[width=7.5cm]{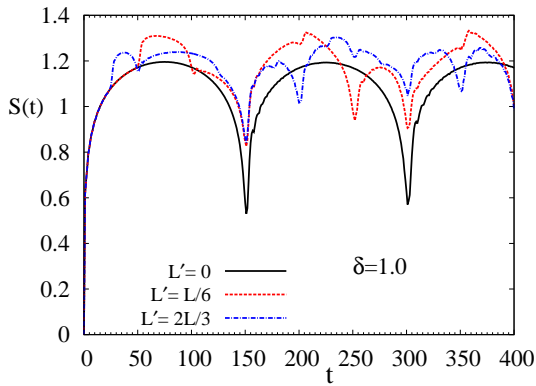}
\end{center}
\caption{
Time evolution of entanglement entropy after 
a local $J$-quench with $h$-quenching at different sites $L'$ for $\de=1.0$. For $L'=L/6=50$,
the deviation from the no $h$-quench case starts at $t'/2=50$. A sharp decrease is observed
at $t=t'=100$ when $QP^2_L$ gets reflected at $L'$ and returns to $L_1=L_A$. There is
a sudden increase in EE after $t=150$ when $QP^2_R$ enters system A whereas $QP^2_L$
is still in system B. We again see a dip at $t''=250$ when $QP^2_L$ and $QP^2_R$
exchange their systems. 
Similarly, for the case $L'=2L/3=200$,
$QP^1_L$ enters system A at $t=t'/2=25$ when the deviation from the single quench
case appears. We see a sudden decrease of EE at $t=t'$ when $QP^2_R$ enters system
A after getting reflected at $L'$ so that both the quasiparticles are in system A.
After $t=150$, once again EE increases as $QP^2_L$ enters system B. A sharp dip is
seen at $t=t''=200$ when $QP^2_L$ and $QP^2_R$ exchange systems.
}
\label{ee_fig1b}
\end{figure}

We show the evolution of LE in Fig.~\ref{le_fig1} for different values of $\de$.
Let us first concentrate on the results of LE with $\delta=0$.
Since LE is the overlap of two wavefunctions which had unity overlap initially, it starts decreasing from one
at $t=0$ till both the quasiparticles
reach the boundary at $t=L/4$ (since $L_1=L_2=L/2$ and $v_{\max}=2$), where it gets reflected. 
Intuitively, during its return path, QPs will undo their effect of dynamics.
Thus, we expect to see a decrease in LE till the reflection of the first
quasiparticle ($i.e.,$ till t=L/4) after which there is 
an increase till the quasiparticle reaches its origin or till $t=L/2(=2L_1/2)$. After this time, the
initially left (right) moving quasiparticle will move to the system 2 (system 1) and eventually come
back to its origin showing a peak at $t=T=L$, 
which is also the time period of the quasiparticles, see Fig. \ref{le_fig1}. 
When the second quench or the h-quench is also performed simultaneously,
we expect to see some structures close to the time scales mentioned before, i.e., at $t_1,t_2,t',t''$
along with the $\delta=0$ structure. 
The numerical results of such double quenches are shown
in Fig. \ref{le_fig1} for various coupling strengths $\delta$ and fixed $L'$.
As expected, the decay in LE is stronger with increasing strength of the local
h-quench or $\delta$.
On the other hand, Fig.~\ref{le_fig2} shows the variation of LE$(t)$ when 
the $h$-quench is performed at different positions of the chain with fixed $\delta=1$ 
demonstrating the above mentioned
time scales more clearly, especially the variation of $t''$ with $L'$. We find good
agreement between the timescales proposed above and the numerics, see the caption for more details.

In conclusion, we find that the dominant peaks are due to 
$QP^2$ at times $t',t'',t_3$ and $t_4$, where $t_3=t_4$ in this case. 
We also note that the peak at $t''$
is a strong peak which may be due to the fact that at this time three different quasiparticles
return to their origin after reflections at various points as discussed below:
(i) $QP^1_R$ after reflection from the right boundary,
(ii)$QP^2_R$ after reflection from right boundary and a second reflection at $L'$
causing it to return to $L_1$
(iii) $QP^2_L$ after reflection at $L'$ and a second reflection at right boundary
resulting to its return to $L_1$.
Finally, a peak is also observed at $t=L$
which is the return time of all the fastest quasiparticles back to their origin
when there is no reflection at $L'$, thus giving us a hint that there may be a transmitted
component of the quasiparticle also. We shall comment more on it after
discussing the results in the ferromagnetic phase.

It is to be noted that the presence of time scale $t_1$ due to $QP^1$ is almost negligible
in these figures, though  we do observe some perturbation at this time. 
On the other hand, it is too early to discard the presence of $QP^1$ as its effect  
is very clearly observed in the evolution of EE as explained in the next paragraph,
thus ruling out the possibility of absence of such quasiparticles. 
We shall try to argue about the
absence of $t_1$ scale in the evolution of LE in section \ref{sec_comparison}.

We now focus on the entanglement entropy as a function of time for the 
above scenario. In this case, another parameter is the size $L_A$
of system A of which we calculate the entanglement entropy with the
remaining system of size $L-L_A$.
Let us first consider the simplest case where $L_1=L_A$, i.e., the location of 
J-quenching also determines the size $L_A$ of system A.
Interestingly, the bipartite EE of two critical transverse Ising chains can also 
detect the response of $h$-quenching (see Fig.~\ref{ee_fig1a}, \ref{ee_fig1b}). 
We see that for $J$ quenching alone or for the single quench, 
the entanglement entropy shows perfect periodic 
oscillations with dips at $t_3=t_4=2L_1/v_{\max}$, also discussed in Ref.\cite{stephan11}
using conformal field theory. 
This can also be explained using the quasiparticle picture. A pair of quasiparticle
will increase the entanglement between the system A and the rest if one of
them is in system A and the other is in B. The quasiparticle pairs are generated at $L_1=L_A=L/2$
and travel in opposite directions resulting to an immediate increase in $S(t)$ for $t>0$. This is 
not the case when $L_1\ne L_A$ and will soon be discussed separately.
As both of them reaches the boundary at $t=L/4$ and gets reflected, $S$ starts decreasing and eventually
shows a dip when $t=L/2$ after which both the quasiparticles belonging to 
a pair exchange their systems and once again EE increases after $t=L/2$. 
At $T=2L/v_{\text{max}}(=L)$,
both the quasiparticles arrive at the starting point and $S$ shows a dip once again after which 
the pattern repeats.
This is also shown in Fig \ref{ee_fig1b} with $L'=0.0$.
If we now perform local $h$-quenching at a general site $L'$ of the spin chain, we observe
that the evolution of EE in double quenches follows the single quench case but accompanied by 
deviations at certain times which can once again be 
explained using the quasiparticle picture. We observe that the double quench case 
follows the single quench case $(\delta=0.0)$ 
till $t=t'/2$ after which there is a sudden deviation or increase from the single quench case.
This is because one of the quasiparticles ($QP^1_R$) produced at $L'$ (which is not present in single quench case) 
enters the system B at this time
whereas the other quasiparticle of the same pair remains in system A. This results
to an extra increase in $S$. On the other hand, at $t=t'$, the $QP^2_L$ reaches back to 
$L_1=L_A$ after reflection at $L'$ where we find a sharp decrease in $S$
as both $QP^2_L$ and $QP^2_R$ are now in system B. 
The EE keeps on decreasing (after a slight increase) till 
$QP^2_R$ enters system A at $t=2(L-L_1)/v_{\max}$ 
as a result of getting reflected from the right boundary,
after which we observe a
sharp increase in EE.
 One more time scale observed corresponds to
$t''=2(L-L')/v_{\max}$ which was also observed in LE. 
At this time, the $QP^2_R$ and $QP^2_L$ return back
to $L_A$ and exchange their systems, as also discussed before with reference to LE.

The main difference between the analysis of LE and EE is that
in the case of LE, we are interested in time scales at which the produced quasiparticles
come back to its origin. In the case of EE, we are interested in the timescales
in which one of the quasiparticles belonging to a pair crosses one system
and goes to the other system. If this crossover results to both the QPs
to be in the same system, then EE decreases, otherwise it increases 
Here, since
we considered the geometry where $L_1=L_2=L_A$, many of the time scales
are hidden.
In the next section, we apply the same ideas to the case when 
$L_1 \neq L_2$ but $L_1=L_A$ along with a special discussion for the 
most general case when
$L_1 \neq L_2 \ne L_A$, and verify the quasiparticle picture proposed.

\begin{figure}[h]
\begin{center}
\begin{overpic}[width=7.5cm]{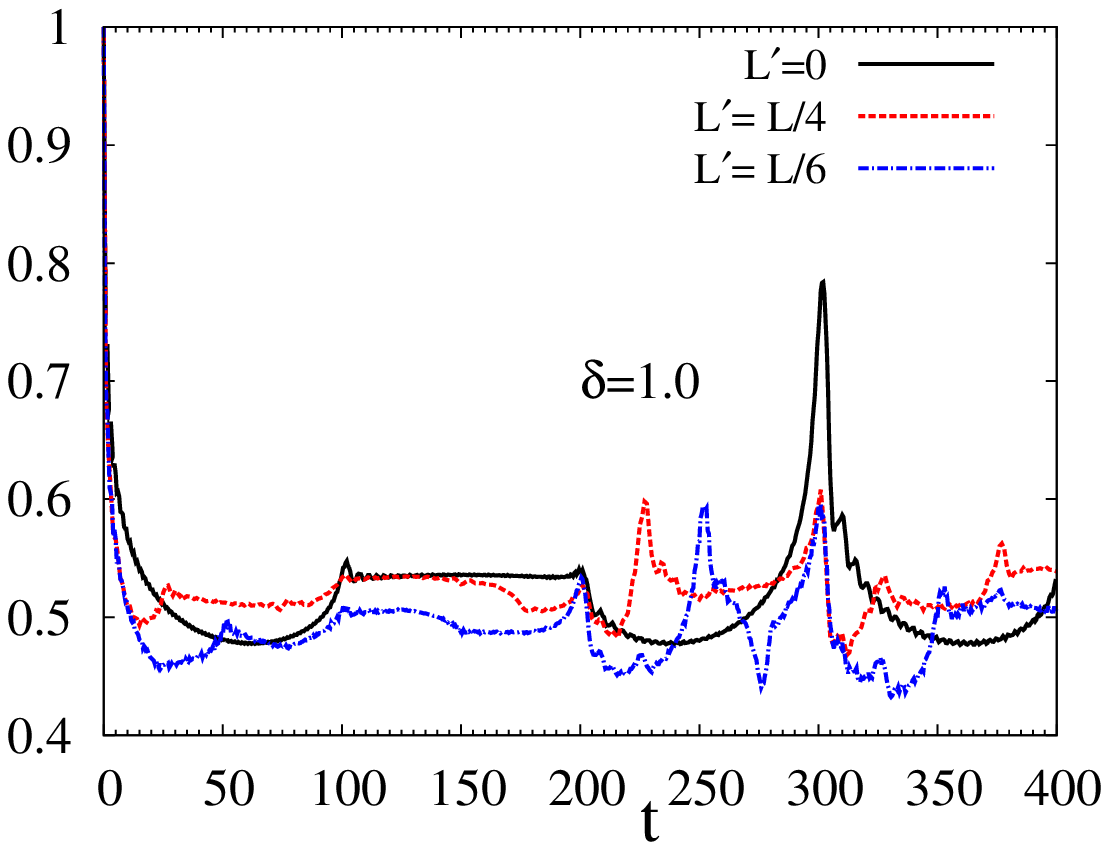}
\put(-4,45){$\mathcal{L}(t)$}
\end{overpic}
\end{center}
\caption{The plot shows LE as a function of time for double quenches 
with $L_1=100,L_A=100$ and $L=300$ and different $L'$. 
The first peak of LE occurs 
at time $t'=2(L_1-L')/v_{\max}$ ($t'=25$ and $50$ for $L'=L/4$ and $L/6$ respectively). 
The other time scales are $t_3=100$, $t_4=200$ and $t''$. 
We note that $t''=225$ and $250$ for $L'=75$ and $50$, respectively.}
\label{le_fig4}
\end{figure}

\subsection{LE and EE for $L_1 \neq L_2$}
\label{sec_l1neql2}
We are now interested in the local quenching of asymmetric spin chain 
($L_1\neq L_2$). 
Let us first study the evolution of Loschmidt echo.
For $J$-quenching alone \cite{stephan11}, the LE shows three time scales 
given by $t_3$, $t_4$ and the time period $T$ discussed in 
section \ref{sec_l1eql2} (see $L'=0$ plot of Fig.~\ref{le_fig4}).
The nature of these plots are discussed in details in Ref. \cite{stephan11}
using conformal field theory. When the transverse field term at $L'$ is 
changed from $h$ to $h+\de$ 
together with J-quenching, we expect to see the following 
additional time scales in parallel with the discussion in the previous
section: $t_1=(2L')/v_{\max}$, 
$t'=2(L_1-L')/v_{\max}$  
, $t''=2(L-L')/v_{\max}$ 
Fig.~\ref{le_fig4} shows LE as a function of time after $J$-quenching at $L_1=L/3$ 
and $h$-quenching at different sites $L'$. All the above mentioned time scales can be
clearly seen in this figure except $t_1$. We shall try to argue for this latter.

\begin{figure}[h]
\begin{center}
\includegraphics[width=7.5cm]{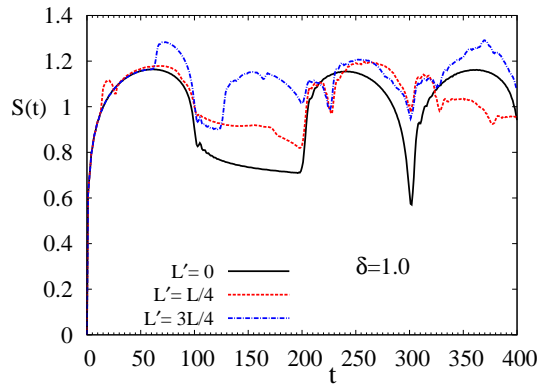}
\end{center}
\caption{Time evolution of entanglement entropy for the same case as in Fig. \ref{le_fig4}
but different $L'$. 
For $L'=L/4$, the deviation from single J-quench case appears at $t'/2=12.5$
whereas at $t'=25$, $QP^2_L$ enters system B after getting reflected at $L'$ where its
other partner $QP^2_R$ is already present. The decrease in EE continues till $QP^2_R$ enters system
A at $t=L_2=200$. We also see a dip at $t=t''=225$ where $QP^2_L$ and $QP^2_R$ exchanges
their systems. 
Similarly, one can argue for the evolution of EE when $L'=3L/4$. The deviation from the single quench
case begins at $t=62.5$. In this case, $QP^2_L$ enters system B at $t=100$ which causes a sharp decrease
at this time. On the other hand, $QP^2_R$ enters system A at $t=125$ resulting to an increase in EE
as its other counterpart is still in B. The natural increase at $t=200$ which is there for 
only J-quenching case can also be observed. This might be due to the fact that any reflection
at $L'$  is not perfect and there is a possibility of getting a transmitted component
of the QP wave, also discussed in sections \ref{le_ee_ferro} and \ref{discussion}. 
The time scales $t''=225$ and $T=300$ are also present.}
\label{ee_fig2}
\end{figure}

Let us now move on to the calculation of EE for the same situation ($L_1\neq L_2$, but $L_1=L_A$).
Fig.~\ref{ee_fig2} shows the time evolution 
of EE after single and double quenching at time $t=0$. One can observe the difference 
in LE and EE between the times $t_3$ and $t_4$ for $\de=0$ case 
(see Fig.~\ref{le_fig4} and Fig.~\ref{ee_fig2}). In this time range LE remains constant. 
On the other hand EE decreases slowly and then starts increasing 
at $t=t_4$ as discussed in Ref. \ct{stephan11}, the increase being due to arrival of
$QP^2_R$ in system A. 
Moving to the double quenches, the discussion is almost same as for the case $L_1=L_2=L/2$ in the previous
section. EE more or less follows the $\delta=0$ case and we see special time scales at $t'/2$, $t'$ and $t''$. 
See caption of Fig. \ref{ee_fig2} for more details.

\begin{figure}[h]
\begin{center}
\includegraphics[width=7.5cm]{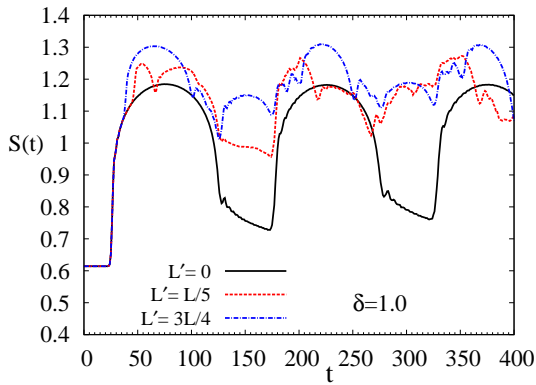}
\end{center}
\caption{Time evolution of entanglement entropy after a local $J$-quenching at $L_1=L/3$ 
 and $h$-quenching at different sites $L'$ with $\de=1.0$.
 Here, the subsystem is of length $L_A=L/2$ which does not coincide with cut resulting to few more
relevant time scales. For $L'=L/5=60$, the first deviation (or increase) from the single quench case 
appears at $t=45$. The EE decreases at $t=t'=65$ when $QP^2_L$ and $QP^2_R$ are in the same subsystem B.
The next increase in EE would be at $t=175$ when $QP^2_R$ enters subsystem A. The split of time scale $t''$,
as discussed in the text, can also be seen with dips at $t''_1=215$ and $t''_2=265$.
The case with $L'=3L/4$ is all the more interesting. The deviation occurs at $t=(L'-L_A)/2=37.5$.
The sharp decrease in this case occurs at $t=100$ which is the time taken by $QP^2_R$ to get reflected at $L'$
and enter system A so that both $QP^2_R$ and $QP^2_L$ are in system A. But at $t=125$, $QP^2_L$ enters system B
after getting reflected from the left boundary resulting to an increase in EE. $t''_1=200$ (due to $QP^2_L$)
and $t''_2=250$ (due to $QP^2_R$). We note extra dips around $t=187$ which seems to be due to 
$QP^1_L$ entering system B after reflection from the left boundary.}
\label{ee_fig3}
\end{figure}

We now consider the most general situation with $L_1 \neq L_2 \neq L_A$ 
and calculate the time evolution of EE. 
Here, $J$-quenching is again performed at $L_1$, but the subsystem is assumed to be of length 
$L_A=L/2$, different from $L_1$. The EE as a function of time after the double quenches 
is shown in Fig.~\ref{ee_fig3}. Let us define $l(=L_A-L_1)$ as the distance between the right end of 
the subsystem A and the site of $J$-quenching. As in other cases, we discuss explicitly
the case with $L'<L_1<L_A$ below, and try to present some other examples through the figures.
For $\delta=0$, EE remains at a constant value for small times and starts increasing at $t=l/v_{\max}$
when $QP^2_R$ hits at the boundary of the two subsystems and enters the subsystem $B$ ~\ct{eisler07}.
Note the contrast between $L_1=L_A$ where $S(t)$ increases immediately and $L_1 \neq L_A$  where
$S(t)$ is constant initially.
The EE shows a sharp decrease at $t=(2L_1+l)/v_{\max}$ when $QP^2_L$ enters system B after getting reflected
from the left boundary. Similar to
the previous case (i.e., for $L_1\neq L_2$ and $L_1=L_A$), there is a 'decay region' between the time range 
$[(2L_1+l)/v_{\max},(L_2+L_B)/v_{\max}]$ where EE decays very slowly, after which there is an increase
as both the QPs are now in different subsystems. 
In this case, since the site for J-quenching
does not coincide with the subsystem size, there are many additional time scales as
discussed below, the appearance of which puts the picture of travelling quasiparticle 
on stronger footing.. 
Let us now come back to the double quenches.
Following the double quenches, EE  more or less follows the single
quench case. The first deviation resulting to an increase in EE 
(similar to the one discussed before) occurs at 
$t=(L_A-L')/v_{\max}$, when $QP^1_R$ enters the subsystem $B$. 
By definition, the time scale $t'$ is the time taken by the $QP^2_L$ to get reflected at $L'$
and come back to system A, which in this particular case is given by $t'=(2(L_1-L')+l)/v_{\max}$. 
On the other hand, the time scale $t''$, defined as the time taken by $QP^2_L$ 
(or $QP^2_R$) to undergo double reflections at $L'$ and one of the boundaries, 
gets divided into two scales. This is because the distance travelled by $QP^2_L$ 
is smaller than $QP^2_R$, which was not the case in our previous discussions where $L_1=L_A$.
Let us define these two timescales by $t''_1=(2(L_1-L')+l+2L_B)/2$ (for $QP^2_L$) and 
$t''_2=t''_1+2l/v_{\max}$ (for $QP^2_R$). All these time scales are clearly shown in Fig. \ref{ee_fig3}.
We would like to point out here that the basic physics related to tracking of QPs remain 
same when we change the position of $L'$, but the formula for these time scales may have to be 
changed as can be seen in the $L'=3L/4$ case
discussed in Fig. \ref{ee_fig3}. Also, the above discussion will be correct if
$(L_A-L')/2 < L_1-L'$, i.e., $QP^1_R$ reaches $L_A$ before $QP^2_L$. In the opposite case also,
one needs to simply apply the same ideas to get the right picture of dynamics.
It is also to be mentioned that we do observe some extra time scales, some of which can be explained
and are discussed in the caption of Fig. \ref{ee_fig3}.

\section{Entanglement Entropy for a ferromganetic chain}
\label{le_ee_ferro}
\begin{figure}[h]
\includegraphics[width=7.5cm]{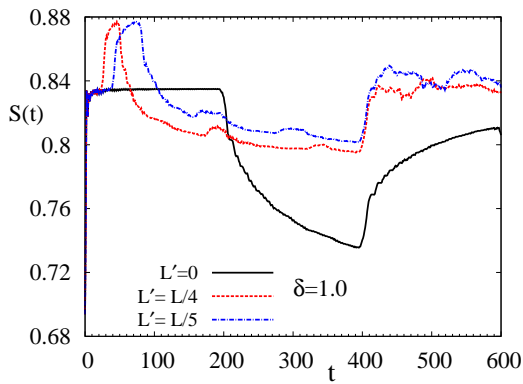}
\caption{The plot shows EE as a function of time for single and double quenches 
when the whole chain is non-critical ($h=0.5$)
with $L_1=100$, $L_A=100$ and $L=300$ and different $L'$.
For $L'=L/4$, the deviation in EE from single quenching case starts at $t=t'/2$ where $t'=50$,
with $v_{\max}=2h=1$. For $t>50$, both $QP^2_L$ and $QP^2_R$ are in system B 
leading to decrease in EE which continues up to $t_4=400$. 
Similarly, for $L'=L/5$ one can find $t'=80$ and the figure shows the expected behavior.
The absence of $t''$ is explained in the text. 
}
\label{ee_fig4}
\end{figure}

\begin{figure}[h]
\includegraphics[width=7.5cm]{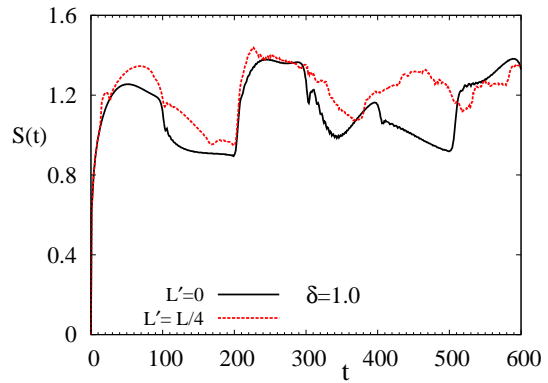}
\caption{
 EE as a function of time for the same situation as in Fig\ref{ee_fig4} but fixing $h$ at $0.99$. In this 
case the value of $v_{\max}$ is $1.98$. This gives $t'=25.25$ for $L'=L/4$. We do see timescales
$t'/2$ and $t'$ along with $t_3$ ($\sim$ 101) and $t_4$ ($\sim$ 202). We also observe a
peak near $t''$ ($\sim$ 227). 
}
\label{ee_fig5}
\end{figure}

In this section, we briefly discuss the evolution of entanglement
entropy when the total spin chain is in the ferromagnetic phase.
Here, we consider local quenching of asymmetric spin chains $(L_1\neq L_2)$ 
with $L_1=L_A$.
We concentrate upon two different cases to calculate EE after single or double quenches 
: one where the total spin chain is deep inside the ferromagnetic phase (see Fig.~\ref{ee_fig4}) 
and the other where the spin chain is close to the critical point (see Fig.~\ref{ee_fig5}).
Let us first consider the spin chain with $h=0.5$ at all sites.
Fig.~\ref{ee_fig4} shows time evolution of EE
after single quench at $L_1$ ($J$ quench) and also after the double quenches, namely,  $h$ quench at $L'$ and $J$ quench at $L_1$.
For the single quench ($L'=0$), EE detects 
$t_3$ and $t_4$ successfully. Note the difference between the critical and ferromagnetic region
for times up to $t_3$. In the critical case, the EE increases at $t=0$ followed by a decrease
which starts around $t_3/2$ when the $QP^2_L$ gets reflected from the boundary, though 
the decrease is sharper at $t_3$.
On the other hand, in the ferromagnetic region, we see a sudden increase in EE
followed by an almost constant EE region up to $t_3$ (no decrease at $t_3/2$) after which
it decreases suddenly. This hints to the fact that in the ferromagnetic region, QPs
are more point like particles, and hence its location can be known precisely. But in the
critical case, these QPs are extended wavepackets as also mentioned in Ref. \cite{blass12},
and hence the reflection at the boundary is felt also at $L_1$. 
Similar to the critical case (see Sec.~\ref{sec_l1neql2}), EE 
decays between times $t_3$ to $t_4$. In this time range the 
fastest moving quasi-particles do not contribute in the EE. Let us now 
discuss the time evolution of EE after double quenches. One can observe clearly 
the time scales $t'/2$ and $t'$ from Fig.~\ref{ee_fig4}. 
Interestingly, EE starts decreasing after $t'$ and it continues up to $t_4$.
The sharp increase in EE at $t_4$ is due to the fact that at $t_4$, $QP^2_R$ 
enters system A whereas $QP^2_L$ is still in system B. It is to be mentioned
that in the ferromagnetic case, $t''$ is not clearly visible, which once again
can be explained due to the point like nature of QPs in the ferromagnetic region.
For $t<t''$, $QP^2_L$ is in system B and $QP^2_R$ is in system A. They exchange their systems
at $t''$, thus contributing to the entropy equally for $t<t''$ and $t>t''$.
On the other hand, the critical case distinguishes between QPs approaching $L_1$
and moving away from $L_1$ due to the finite extent of QP.

We now move to explain the quenching results when the final system remains close to the 
quantum critical point. The evolution of EE as a function of time following single and double quenches
is shown in Fig.~\ref{ee_fig5} for $h=0.99$. In this case also, we observe a sudden increase/deviation
from the single quench case at $t'/2$ followed by a sharp decrease at $t'$. We also observe an increase
immediately after $t'$ which is different from the $h=0.5$ case and similar to the critical case.
This may be because the QP which is now more like an extended object with extended wavefunction is only partially 
reflected at $L'$  as compared to localized QP deep inside the ferromagnetic phase
having less wavelike properties.
Rest of the discussion is the same in this case and discussed in details in the caption of Fig. \ref{ee_fig5}.

\section{Comparison between $QP^1$ and $QP^2$}
\label{sec_comparison}

\begin{figure}[h]
\includegraphics[height=3.0in,angle=-90]{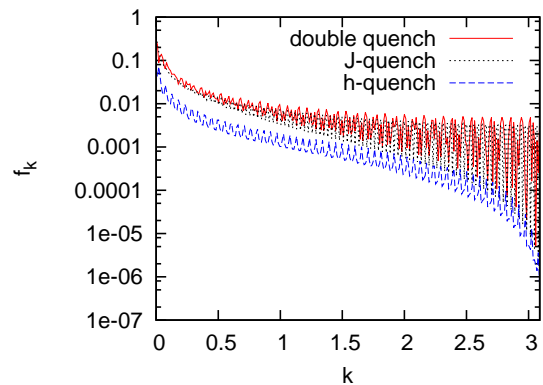}
\caption{Variation of $f_k$ with $k$ for a critical chain.
The dashed line or the lowest curve corresponds to the h-quench case
which clearly is an order of magnitude smaller than the J-quench alone. 
Similar behavior is also observed for a ferromagnetic chain.}
\label{fig_fpcomp}
\end{figure}

In this section we try to argue why the effect of $QP^2$ is stronger than $QP^1$.
As discussed before, the initial state, which is no longer the ground state of the final Hamiltonian, 
is a source of quasiparticles. Also, since the perturbation studied in this paper is local and very small,
the quasiparticles are produced at the site of perturbation only, i.e., at $L'$ and $L_1$.
These quasiparticles have energies given by the eigenvalues of the final Hamiltonian.
Let $k$ identifies the quasiparticle $\eta_k$ having eigen energy $\Lambda_k$.
Quasiparticles of energy $\Lambda_k$ are produced at the perturbation site with probability $f_k$.
Numerically, one can obtain $f_k$ by calculating the expectation value
$$f_k=\langle \psi_i|\eta_k^\dagger \eta_k |\psi_i \rangle$$
which is proportional to the number of quasiparticles $\eta_k$ present in the initial state $|\psi_i\rangle$.
This expression can be written in terms of $\Phi_k$, $\Psi_k$ of the final Hamiltonian 
and the matrix $G^i$ (see Appendix \ref{appendixb}) with respect to the initial Hamiltonian.
A comparison of $f_k$ for the h-quench alone ($h$ changed from $1$ to $2$ at $L'$), $J$-quench alone (0 to 1 at $L_1$) 
and both quenches together is shown in 
Fig. \ref{fig_fpcomp}. Clearly, quasiparticle creation probability is an order of magnitude
higher in case of $J$-quench alone when compared to $h$-quench. This hints to the fact that the 
$J$-quench is the main source of quasiparticle production and hence our numerical results
are dominated by the dynamics of $QP^2$.

\section{Conclusions and Discussions}
\label{discussion}
In this paper, we studied the effect of two simultaneous local quenches
in an otherwise uniform transverse Ising chain of length $L$ with open boundary conditions.
Initially, the system is prepared in the ground state of the transverse Ising chain
having a uniform transverse field $h$ and
the interaction strength set to unity at all sites except $J_{L_1}$
and $J_L$ where it is zero. 
The first quench corresponds to sudden increase of $J_{L_1}$ from zero
to $1$, and the second quench  
involves the sudden change of the transverse field from $h$ to $h+\delta$
at site $L'$.  
We argued that the sites of the two local quenches 
are source of quasiparticle production as there is a local increase in 
energy due to the quenches. These QPs are wavepackets of low lying
excitations of the final Hamiltonian.
As discussed in Refs \cite{rieger11,blass12}, the QPs are localized in the ferromagnetic
region and behave more like classical particles, whereas they are extended objects/wavepackets 
as the critical point is approached.

We numerically studied the evolution of Loschmidt echo and
entanglement entropy after the double quenches and explained the evolution
using the quasiparticle picture. 
The envelope of the curve is dictated by the fastest moving quasiparticles.
We showed taking examples that most of the 
timescales can be explained
using the propagation of quasiparticles.
The most interesting phenomena that is observed numerically is the
partial or full reflection of the quasiparticles at $L'$, the site of $h-$quench.
Only if we include such a phenomena that we can explain certain numerically observed
time scales.
As mentioned in details in the paper, the most relevant time scales are $t'$, $t_3$, $t_4$
and $t''$ (for $L_1=L_A$) which are all due to the $QP^2$ pair, or the pair 
produced at the J-quenching site. The presence of the other set of quasiparticles, namely, $QP^1$ is
clearly seen in the evolution of EE. We have shown that the probability of quasiparticles
produced due to h-quench is roughly an order of magnitude smaller than the J-quench which could be the 
reason for stronger effect of $QP^2$ in the evolution of LE and EE. 

The double quenches deep inside the ferromagnetic phase can be
very nicely described by the point like quasiparticles where all the time
scales are sharply observed. We have contrasted this ferromagnetic case
with the double quenches in the critical phase and proposed the reasons
for their differences.
It seems that the reflection of $QP^2$ at $L'$ in a critical chain
is only partial having a transmitted component also. This can be attributed to
extended wavepacket nature of quasiparticles at the critical point. 
One can then explain the decrease of EE
at $t=t_3$, slight increase of EE for $t>t'$ after the sharp decrease at 
$t'$, and the dip at $t=T$. 

Our main aim in this paper is to study the dynamical evolution of LE and EE
after double quenches and see if one can explain the behavior, atleast qualitatively, 
using propagation of quasiparticles. We have demonstrated here that this indeed is possible.
Though we can not propose a general formula for all the timescales involved as it depends
on which quasiparticle arrives at the subsystem first, which in turn depends on
the location of $L'$, $L_1$ and $L_A$, but the basic idea gives us the right picture.
We have checked this for other cases also which are not presented in this paper.
The quasiparticle picture does explain many features, if not all, of the dynamical evolution
of LE and EE that occur in double quenches studied here. 
We have provided some arguments for the immediate increase of EE after $t'$
for a critical chain which is related to the extended nature of the 
QPs at the critical point.

{\bf {Acknowledgements}}
UD acknowledges funding from DST-INSPIRE Faculty fellowship (IFA12-PH-45) by DST, Govt. of India.
AR and UD sincerely thank Amit Dutta for fruitful discussions and the hospitality
of IIT Kanpur where some parts of this work were done. AR acknowledges Bikas K Chakrabarti 
for useful comments.
\appendix
\section{Loschmidt echo for a general quadratic fermionic system}
\label{appendixa}
Here we shall discuss the method for evaluating the time evolution of LE in real space 
for a general quadratic fermionic system~\ct{rossini07}.
We rewrite the Hamiltonian in Eq.~(\ref{ham_quadratic1}) in the following form
\beq
H = \frac{1}{2} {\bf C^\dagger \, \mathcal{H} \, C},
\label{ham_quadratic2}
\eeq
where ${\bf C^\dagger} = \left( c_1^\dagger, \ldots c_L^\dagger, \,c_1, \ldots c_L \right)$ 
and ${\bf \mathcal{H}} = \sigma^z \otimes {\bf A} + i \sigma^y \otimes {\bf B}$.

The Loschmidt echo, defined in Eq.~(\ref{le1}) can be evaluated in this case by the 
following relation~\ct{levitov96,rossini07}
\beq
\mathcal{L}(t)=|\langle\psi(0)|e^{-i t H_f}|\psi(0)\rangle|
=|\mathrm{det}(1-\bf{R}+\bf{R}e^{-i \bf{\mathcal{H}_f} t})|,
\label{le2}
\eeq
where $H_f$ is the final Hamiltonian after double quenches. $\bf{R}$ is the $2 L \times 2 L$ correlation matrix whose elements 
are two-point correlation functions of fermionic operators $R_{ij}=\langle\psi(0)|C_i^\dagger C_j|\psi(0)\rangle$, 
where $|\psi(0)\rangle$ is the ground state of the initial Hamiltonian $H_i$. Following some mathematical steps we can 
the matrix $\bf{R}$ in terms of $\bf{g}$ and $\bf{h}$ matrices (see discussion around Eq. \ref{eq_gh})
\beq
{\bf R} = \left( \begin{array}{cc}
{\bf h^i} \; {\bf (h^i)^T} &
\quad
{\bf h^i} \; {\bf (g^i)^T} \\
{\bf g^i} \; {\bf (h^i)^T} &
\quad
{\bf g^i} \; {\bf (g^i)^T}
\end{array} \right). 
\label{corr_fermion}
\eeq
Here, the $\bf{g}^i$ and $\bf{h}^i$ matrices are calculated relative to the initial Hamiltonian $H_i$.
For more details, see Ref. \cite{rossini07}

\section{Time dependant entanglement entropy for a general quadratic fermionic system}
\label{appendixb}
We outline here steps for calculating the time evolution of EE after sudden quenches~\cite{igloi00,igloi09}. As discussed in the text, 
our model is reduced in quadratic form using spinless fermions. Let us define two Clifford operators which are also 
related to the Majorana fermion operators $a_{2i-1}$ and $a_{2i}$ in the following way,
\bea
\mathcal{A}_i=c_i^{\dag}+c_i= a_{2i-1}
\hspace{2mm}\text{and}
\hspace{2mm}\mathcal{B}_i=c_i^{\dag}-c_i= i a_{2i}.
\label{clifford1}
\eea
The operators $\mathcal{A}$ and $\mathcal{B}$ can be written in terms of the free-fermion operators $\eta_k$

\beq
\mathcal{A}_i=\sum_{k=1}^L \Phi_k(i)(\eta_k^\dag+\eta_k),
\hspace{2mm}\mathcal{B}_i=\sum_{k=1}^L \Psi_k(i)(\eta_k^\dag-\eta_k).
\label{clifford2}
\eeq

The time evolution of these operators are obtained from the time dependence of fermionic operators,
i.e., $\eta_k(t)=e^{-it\Lambda_k}\eta_k$, where $\eta_k$ and $\Lambda_k$ are 
quasiparticles and eigen energies corresponding to the final Hamiltonian $H_f$. This is given by

\bea
\mathcal{A}_i(t)=\sum_{j=1}^L\left[ \langle \mathcal{A}_i \mathcal{A}_j \rangle_t \mathcal{A}_j +
 \langle \mathcal{A}_i \mathcal{B}_j \rangle_t \mathcal{B}_j\right], \cr
\mathcal{B}_i(t)=\sum_{j=1}^L\left[ \langle \mathcal{B}_i \mathcal{A}_j \rangle_t \mathcal{A}_j +
 \langle \mathcal{B}_i B_j \rangle_t \mathcal{B}_j\right],
\label{time_cliff1}
\eea
where the time-dependent contractions are 

\bea
\langle \mathcal{A}_i \mathcal{A}_j \rangle_t&=&\sum_{k=1}^L \cos( \Lambda_k t) \Phi_k(i) \Phi_k(j)\; ,\nonumber\\
\langle \mathcal{A}_i \mathcal{B}_j \rangle_t&=&\langle B_j A_i \rangle_t=i \sum_{k=1}^L \sin (\Lambda_k t) \Phi_k(i) \Psi_k(j)\; ,\nonumber\\
\langle \mathcal{B}_i \mathcal{B}_j \rangle_t&=&\sum_{k=1}^L \cos (\Lambda_k t) \Psi_k(i) \Psi_k(j)\; .
\label{time_cliff2}
\eea

The total system is divided into two subsystems $A$ and $B$ of length $L_A$ and $L_B$ respectively. 
To evaluate EE between these two subsystems after the double quench,
we have to calculate the reduced density matrix $\rho_A$ of the subsystem $A$.
This can be reconstructed 
from the $2L_A\times2L_A$ correlation matrix of the Majorana operators

\beq
\langle \psi_i|a_m(t)a_n(t)|\psi_i \rangle = \de_{m,n} + \imath (\Gamma^A)_{mn},
\label{corr_majorana}
\eeq
where $m,n=1,2,3,\cdots,2L_A$ and $|\psi_i\rangle$ is the initial ground state. 
The matrix $\Gamma^A$ is an antisymmetric matrix which can be brought into the block-diagonal 
form by an orthogonal matrix say, $V$. Therefore the eigenvalues of $\Gamma^A$ are purely imaginary of the 
form $\pm \nu_l$ with $l=1,2,\cdots,L_A$. This can be used to write the reduced density matrix as a direct product 
of $L_A$ uncorrelated modes $\rho_A=\otimes_{l=1}^{L_A}\varrho_l$, where each $\varrho_l$ has eigenvalues 
$(1\pm\nu_l)/2$. Thus the bipartite EE for $\rho_A$ is the sum of entropies of $L_A$ uncorrelated modes
given by
\beq
S_L(L_A)=-\sum_{l=1}^{L_A} \left(\frac{1+\nu_l}{2} \log \frac{1+\nu_l}{2}
+\frac{1-\nu_l}{2} \log \frac{1-\nu_l}{2}\right).
\label{entropy}
\eeq

The time-dependent expectation values in Eq.~(\ref{corr_majorana}) are calculated using Eq.~(\ref{time_cliff1}) as
\bea
\langle a_{2l}(t)a_{2m}(t)\rangle=-\langle \mathcal{B}_l(t)\mathcal{B}_m(t)\rangle,\cr
\langle a_{2l-1}(t)a_{2m-1}(t)\rangle=-\langle \mathcal{A}_l(t)\mathcal{A}_m(t)\rangle,\cr
\langle a_{2l}(t)a_{2m-1}(t)\rangle=-i \langle \mathcal{B}_l(t)\mathcal{A}_m(t)\rangle,\cr
\langle a_{2l-1}(t)a_{2m}(t)\rangle=-i \langle \mathcal{A}_l(t)\mathcal{B}_m(t)\rangle.
\eea
Thus we shall get the time-dependent correlation matrix $\Gamma^A$ and for each time we can obtain EE from Eq.~(\ref{entropy}). 
The elements of the matrix $\Gamma^A$ are given by

\bea
\Gamma^A_{2l-1,2m-1}&=&-\imath \sum_{k_1,k_2} G^{i}_{k_1k_2} \langle \mathcal{A}_l \mathcal{B}_{k_1} \rangle_t
\langle \mathcal{A}_m \mathcal{A}_{k_2} \rangle_t \nonumber\\
&+&\imath \sum_{k_1,k_2} G^{i}_{k_2k_1} \langle \mathcal{A}_l \mathcal{A}_{k_1}\rangle_t \langle \mathcal{A}_m \mathcal{B}_{k_2} \rangle_t \non \\ 
\Gamma^A_{2l-1,2m}&=& \sum_{k_1,k_2} G^{i}_{k_2k_1} \langle \mathcal{A}_l \mathcal{A}_{k_1} \rangle_t
\langle \mathcal{B}_m \mathcal{B}_{k_2} \rangle_t\nonumber \\
&-&\sum_{k_1,k_2} G^{i}_{k_1k_2} \langle \mathcal{A}_l \mathcal{B}_{k_1} \rangle_t \langle \mathcal{B}_m \mathcal{A}_{k_2}\rangle_t \non \\
\Gamma^A_{2l,2m-1}&=& -\sum_{k_1k_2} G^{i}_{k_1k_2} \langle \mathcal{B}_l \mathcal{B}_{k_1} \rangle_t
\langle \mathcal{A}_m \mathcal{A}_{k_2} \rangle_t\nonumber \\
&+&\sum_{k_1,k_2} G^{i}_{k_2k_1} \langle \mathcal{B}_l \mathcal{A}_{k_1} \rangle_t \langle \mathcal{A}_m \mathcal{B}_{k_2}\rangle_t \non \\
\Gamma^A_{2l,2m}&=&-\imath \sum_{k_1,k_2} G^{i}_{k_2k_1} \langle \mathcal{B}_l \mathcal{A}_{k_1} \rangle_t
\langle \mathcal{B}_m \mathcal{B}_{k_2} \rangle_t\nonumber \\
&+&\imath \sum_{k_1,k_2} G^{i}_{k_1k_2} \langle \mathcal{B}_l \mathcal{B}_{k_1}\rangle_t \langle \mathcal{B}_m \mathcal{A}_{k_2} \rangle_t. 
\eea
where $G^i_{k_1k_2}=-\sum_k \Psi_k^i(k_1) \Phi_k^i(k_2)$ is the equilibrium correlation function which is 
calculated with the initial Hamiltonian $H_i$.

\end{document}